\documentclass[aip,reprint]{revtex4-1}

\usepackage{graphicx}
\usepackage{dcolumn}
\usepackage{bm}

\usepackage[utf8]{inputenc}
\usepackage[T1]{fontenc}
\usepackage{mathptmx}
\usepackage{etoolbox}
\usepackage{subcaption}
\usepackage{svg}
\usepackage{siunitx}
\usepackage[hidelinks]{hyperref}
\usepackage[version=4]{mhchem}
\usepackage{tabularx, booktabs}
\usepackage{arydshln}
\usepackage{ulem}
\usepackage{textalpha}

\DeclareSIUnit{\photons}{photons}
\DeclareSIUnit[per-mode = symbol]{\pps}{\photons \per \s}
\DeclareUnicodeCharacter{202F}{\,}
\DeclareUnicodeCharacter{2009}{\,}
\DeclareUnicodeCharacter{2032}{\,}

\draft 

\begin{document}


\title{High-resolution MHz time- and angle-resolved photoemission spectroscopy based on a tunable vacuum ultraviolet source} 



\author{Lukas Hellbrück}
\affiliation{Institute of Physics, Laboratory for Ultrafast Microscopy and Electron Scattering (LUMES), École Polytechnique Fédérale de Lausanne (EPFL), CH-1015 Lausanne}
\affiliation{Institute of Physics, Laboratory for Quantum Magnetism (LQM), École Polytechnique Fédérale de Lausanne (EPFL), CH-1015 Lausanne}
\affiliation{Lausanne Centre for Ultrafast Science (LACUS), École Polytechnique Fédérale de Lausanne (EPFL), CH-1015 Lausanne}
\author{Michele Puppin}
\affiliation{Lausanne Centre for Ultrafast Science (LACUS), École Polytechnique Fédérale de Lausanne (EPFL), CH-1015 Lausanne}
\author{Fei Guo}
\affiliation{Institute of Physics, Spin Orbin Interaction Spectroscopy (SOIS), École Polytechnique Fédérale de Lausanne (EPFL), CH-1015 Lausanne}
\author{Daniel D. Hickstein}
\affiliation{Kapteyn–Murnane Laboratories, 4775 Walnut Street Suite 102, Boulder, Colorado 80301, USA}
\affiliation{Octave Photonics, 325 W South Boulder Rd Suite B1, Louisville, CO 80027}
\author{Siham Benhabib}
\affiliation{Institute of Physics, Laboratory for Ultrafast Microscopy and Electron Scattering (LUMES), École Polytechnique Fédérale de Lausanne (EPFL), CH-1015 Lausanne}
\affiliation{Laboratoire de physique des Solides, Phénomènes ultrarapides lumière-solides (PULS), Université Paris-Saclay, FR-91191 Gif-sur-Yvette}
\author{Marco Grioni}
\affiliation{Laboratory of Electron Spectroscopy (LSE), École Polytechnique Fédérale de Lausanne (EPFL), CH-1015 Lausanne}
\author{J. Hugo Dil}
\affiliation{Institute of Physics, Spin Orbin Interaction Spectroscopy (SOIS), École Polytechnique Fédérale de Lausanne (EPFL), CH-1015 Lausanne}
\author{Thomas LaGrange}
\affiliation{Institute of Physics, Laboratory for Ultrafast Microscopy and Electron Scattering (LUMES), École Polytechnique Fédérale de Lausanne (EPFL), CH-1015 Lausanne}
\author{Henrik M. R{\o}nnow}
\affiliation{Institute of Physics, Laboratory for Quantum Magnetism (LQM), École Polytechnique Fédérale de Lausanne (EPFL), CH-1015 Lausanne}
\author{Fabrizio Carbone}
\email[]{fabrizio.carbone@epfl.ch}
\affiliation{Institute of Physics, Laboratory for Ultrafast Microscopy and Electron Scattering (LUMES), École Polytechnique Fédérale de Lausanne (EPFL), CH-1015 Lausanne}
\affiliation{Lausanne Centre for Ultrafast Science (LACUS), École Polytechnique Fédérale de Lausanne (EPFL), CH-1015 Lausanne}


\date{\today}

\begin{abstract}
Time and angle-resolved photoemission spectroscopy (trARPES) allows direct mapping of the electronic band structure and its dynamic response on femtosecond timescales. Here, we present a new ARPES system, powered by a new fiber-based femtosecond light source in the vacuum ultraviolet (VUV) range, accessing the complete first Brillouin zone for most materials. We present trARPES data on \ce{Au(111)}, polycrystalline Au, \ce{Bi2Se3} and \ce{TaTe2}, demonstrating an energy resolution of \SI{21}{\milli\electronvolt} with a time resolution of $<$\SI{360}{\femto\s}, at a high repetition rate of \SI{1}{\mega\hertz}. The system is integrated with an extreme ultraviolet (EUV) high harmonic generation (HHG) beamline, enabling excellent tunability of the time-bandwidth resolution.
\end{abstract}


\maketitle 

\section{Introduction}

Understanding quantum materials and their emergent phenomena is one of the biggest challenges in condensed matter physics \cite{giustino_2021_2020}.
Precise knowledge of the electronic properties of materials is required to understand diverse and complex phenomena such as high-temperature superconductivity \cite{keimer_quantum_2015, scalapino_common_2012, michon_thermodynamic_2019, cao_unconventional_2018}, topological properties \cite{hasan_colloquium_2010, schuffelgen_selective_2019, deng_quantum_2020, liu_robust_2020} or charge density waves (CDW) \cite{chen_charge_2016, lian_unveiling_2018}.
Additionally, access to the temporal degree of freedom enables observations of hidden- or meta-stable states \cite{giannetti_ultrafast_2016, nova_metastable_2019, basov_towards_2017}, and allows the coherent control of such exotic phases\cite{zhang_dynamics_2014, rudner_band_2020, tengdin_imaging_2022}.
A prominent example of this is the recently discovered photo-induced superconductivity \cite{fausti_light-induced_2011, mitrano_possible_2016}.
The complexity of these materials results from many interacting degrees of freedom responsible for the emergence of such effects.
Microscopic couplings encompass multiple time-energy scales, making studies of correlated materials experimentally challenging.
For example, quasiparticle scattering rates in excited CDWs are on the order of a few tens of \si{\femto\s}\cite{maklar_coherent_2022}, whereas phonon-mediated scattering typically occurs on the \si{\pico\s} timescale.
Time- and angle-resolved photoemission spectroscopy (trARPES) has been established as one of the fundamental tools for studying quantum materials\cite{wang_observation_2013, rohwer_collapse_2011, schmitt_transient_2008, mor_ultrafast_2017, boschini_collapse_2018, parham_ultrafast_2017, smallwood_tracking_2012}.
In such experiments, a femtosecond laser pulse is used to generate an excitation whose effect can then be tracked in the electronic band structure with momentum resolution as a function of time, on femtosecond timescales.

Using tunable light sources from \SIrange{6}{100}{\electronvolt} is desirable for addressing specific electronic states in many materials, with the time and energy resolution tailored to the particular effect and dynamics under investigation.
A high signal-to-noise ratio is needed while simultaneously avoiding space-charge effects, leading to artifacts and broadened spectra.
Due to current technological limitations, a single light source capable of covering all these requirements is lacking.
Here, we present a new trARPES system utilizing a tunable light source operating in the Vacuum Ultraviolet (VUV) regime, working in the range of \SIrange{7.2}{10.8}{\electronvolt} at high repetition rates, exceeding typical crystal-based trARPES sources in terms of photon energy and improving on high harmonic generation (HHG) systems in terms of energy resolution.
The light source is based on the cascaded generation of even and odd harmonics of a ytterbium laser in a hollow core capillary fiber.
Isolated harmonics are selected in a near-normal incidence grating monochromator, combining energy tunability up to \SI{10.8}{\electronvolt}, a narrow spectral bandwidth \SI{17}{\milli\electronvolt}, and a monochromatic flux exceeding \SI{1e10}{\pps}.
We characterize the setup by performing trARPES experiments on several materials and demonstrate a temporal resolution as low as \SI{360}{\femto\s}.
An additional benefit of the light source is the high stability due to the fiber-based front end, significantly increasing the ease of use by reducing the laborious daily realignments.
This alignment stability further leads to a stable VUV flux over multiple days, even when switching between different harmonics.
Due to this reproducibility, the light source can be used under various conditions with little to no setup time when changing the characteristics of the source.
The reproducibility also allowed us to integrate the new light source into an existing HHG beamline, thus enabling full flexibility in the choice of photon energies and experimental time-energy scales, with the possibility of switching between the two in a matter of seconds.

\section{Light source}
trARPES experiments require a high photon energy, femtosecond light source, capable of overcoming the sample's work function (\SIrange{4}{5}{\electronvolt}).
Higher photon energies allow studies of electrons carrying higher parallel momentum $k_\parallel$.
For a typical Brillouin zone (BZ) size of \SI{1.64}{\angstrom\tothe{-1}} as in the case of  \ce{Bi2Sr2CaCu2O8_{+\delta}} (Bi2212), VUV light pulses of \SI{10.8}{\electronvolt} are sufficient to access the full first Brillouin zone  (FBZ) in ARPES.
Energy tunability is also desirable to address specific electronic states, enabling studies of quantum materials with stronger 3D character, by studying their out-of-plane momentum component $k_\perp$\cite{bao_ultrafast_2022}.
Photoelectron spectroscopies are susceptible to space charge effects, limiting the energy resolution, when multiple photoelectrons are emitted within a short light pulse.
High pulse repetition rates ($>\SI{100}{\kilo\hertz}$) are therefore desirable to minimize space charge effects.
However, it must be noted that this comes at the cost of limited pump excitation fluence in pump-probe experiments, due to the higher thermal load on the sample.
Limiting the fluence makes observing far out-of-equilibrium effects highly challenging for high repetition rate systems, better suited to exploring the weak perturbation regime.

The most commonly used light sources rely on the harmonic generation of a variety of near-infrared femtosecond lasers in nonlinear crystals such as $\beta$-barium borate (BBO) or potassium beryllium fluoroborate (KBBF) crystals.
With these crystals, it is possible to create harmonics with an energy of up to \SI{6.05}{\electronvolt} for BBOs and \SI{7.56}{\electronvolt} for KBBF crystals \cite{liu_development_2008}.
Crystal-based trARPES systems typically operate at time resolutions >\SI{50}{\femto\s} or even higher to not compromise energy resolution, typically <\SI{70}{\milli\electronvolt} \cite{faure_full_2012, sobota_ultrafast_2012, andres_separating_2015, huang_high-resolution_2022}.
An overview of crystal-based trARPES sources was recently given by Gauthier et al. \cite{gauthier_tuning_2020}.
High conversion efficiency in solids allows for light sources based on high repetition rate laser systems of up to \SI{80}{\mega\hertz} with lower peak intensity, providing high signal-to-noise ratios and avoiding detrimental space charge.
A drawback is the low photon energy, which limits the accessed range of parallel momenta $k_\parallel$, smaller than most materials' entire FBZ.

Photon energies in the VUV can be produced by nonlinear interaction in gases, particularly by third or higher-order harmonic generation of ultraviolet pulses, or by non-perturbative HHG.
The former proceeds typically in two conversion steps, the fundamental is first frequency doubled or tripled in non-linear crystals.
After that, a third or higher-order nonlinear process in a gas is used to generate VUV radiation.
Efficient phase matching is possible in the anomalous frequency dispersion region of the gaseous medium\cite{mahon_third-harmonic_1979}, which is achieved only for specific wavelengths.
Examples are the generation of $9\omega$ for ytterbium or neodymium lasers\cite{peli_time-resolved_2020, lee_high_2020, kawaguchi_time-_2023} or $8\omega$ for Titanium sapphire systems\cite{karlsson_system_1996}.
This process requires higher peak intensities compared to the conversion in crystals but can still be operated at high repetition rates with relatively long driving pulses, ensuring a narrow VUV spectral linewidth ($<\SI{1}{\milli\electronvolt}$\cite{berntsen_experimental_2011}).
The time and energy resolution of ultrafast sources based on third harmonic generation (THG) in gases depends on the chosen design trade-offs.
Examples for the resolutions are the setup proposed by Simone Peli et al. $\Delta E=\SI{26}{\milli\electronvolt}$ and $\Delta t=\SI{700}{\femto\s}$ at \SIrange{1}{4}{\mega\hertz}\cite{peli_time-resolved_2020}, Changmin Lee et al. $\Delta E=\SI{16}{\milli\electronvolt}$ and $\Delta t=\SI{250}{\femto\s}$ at \SIrange{100}{250}{\kilo\hertz}\cite{lee_high_2020} and Kaishu Kawaguchi et al. $\Delta E=\SI{26}{\milli\electronvolt}$ and $\Delta t=\SI{360}{\femto\s}$ at \SI{1}{\mega\hertz}\cite{kawaguchi_time-_2023}.
These results compare well with the current work, where we demonstrate an energy resolution of \SI{21}{\milli\electronvolt} with a time resolution of \SI{360}{\femto\s} at repetition-rates between \SIlist{0.5;2}{\mega\hertz}.
One downside to this technique is the lack of tunability of the light source, limited to odd multiples of the driving UV photon energy in inversion-symmetric media.

Tabletop-based HHG can reach even higher photon energies, providing photons up to the soft X-ray regime\cite{gagnon_soft_2007}.
In such a process, the fundamental light pulse ionizes the atoms of the gas and accelerates the created free electrons in its light field.
These electrons may recombine with the parent atoms, emitting higher energy light in the form of odd harmonics of the fundamental light pulse.
The non-perturbative process results in a comb of harmonics of similar intensity, allowing tunability with a harmonic spacing of $2\omega$. 
The conversion efficiency is significantly reduced when compared to the previous cases, and the process requires shorter pulses with higher peak intensities from the driving laser.
HHG systems, therefore, often operate in a low repetition rate regime <\SI{20}{\kilo\hertz} \cite{frietsch_high-order_2013} but higher repetition rate systems are becoming widespread \cite{guo_narrow_2022, cucini_coherent_2020, puppin_time-_2019, mills_cavity-enhanced_2019, sie_time-resolved_2019, corder_ultrafast_2018, chiang_efficient_2015, chen_time-resolved_2023}.
HHG sources achieve pulses with a shorter time duration of <\SI{20}{\femto\s} but the harmonic spectral width is typically larger than in conventional (perturbative) nonlinear processes, limited by the complex phase matching.
This results in a typical resolution of >\SI{100}{\milli\electronvolt}\cite{puppin_time-_2019, roth_photocarrier-induced_2019, eich_time-_2014, wallauer_intervalley_2016}. 
Furthermore, low repetition rate systems also suffer from space charge broadening, limiting the usable flux for a given experiment.

\begin{figure*}
\includegraphics[width=2\columnwidth]{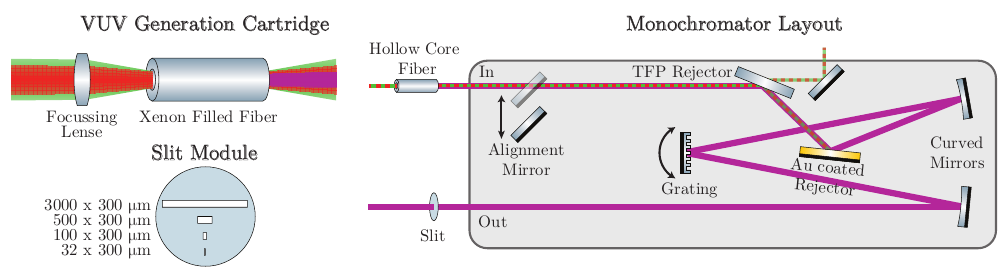}
\caption{Schematic of the Czerney-Turner type monochromator with near normal-incidence (NI) grating. The left side shows a schematic zoom of the HCPCF and in-coupling process and the slit module with the four differently sized exit slits.
\label{fig:Mono_Sketch}}
\end{figure*}

This paper presents a different approach for obtaining photon energies in the VUV regime, which uses a so-called highly cascaded harmonic generation (HCHG) process, creating photons with energies $\leq\SI{16.8}{\electronvolt}$\cite{couch_ultrafast_2020} and the range from \SIrange{7.2}{10.8}{\electronvolt} efficiently usable for ARPES when paired with a near normal-incidence monochromator.
The spectral VUV lines are created in a series of four-wave mixing processes between the fundamental and the second harmonic of an ultrafast ytterbium laser, in a negative curvature hollow-core photonic crystal fiber (HCPCF) filled with Xenon gas.
The HCHG process results in both even and odd harmonics of the fundamental laser wavelength, providing greater tunability than other sources. 
The Xenon gas pressure inside the fiber can be used to optimize phase-matching conditions for the different harmonics\cite{couch_ultrafast_2020}.

The laser source ("Y-Fi", KMLabs Inc.) consists of a fiber-based ytterbium oscillator and amplification front end, followed by the VUV generation stage and a monochromator to select a single harmonic.
The repetition rate can be tuned between \SI{500}{\kilo\hertz}\textendash\SI{2}{\mega\hertz}.
After amplification, pulses of central wavelength $\lambda=\SI{1030}{\nano\meter}$ with an energy of \SI{10}{\micro\joule} are measured after a Treacy grating compressor.
The pulse duration was measured by frequency-resolved optical gating (FROG) and resulted in a typical value of $\sim\SI{250}{\femto\s}$, full width at half maximum (FWHM).
The fundamental beam of frequency $\omega$ is split by a waveplate and polarizing beam splitter to control the amplifier output power.
A pulse with an energy of \SI{3.5}{\micro\joule} at \SI{1}{\mega\hertz} repetition rate is utilized for VUV generation, while the remaining \SI{6.5}{\micro\joule} can be used as a pump pulse in pump-probe experiments.
For the VUV generation, approximately half of the fundamental frequency is doubled in a BBO nonlinear crystal.
The conversion is achieved with a second polarizing beamsplitter, controlling the relative intensity between the $\omega$ and $2\omega$ components to achieve optimal wave-mixing efficiency for the following VUV generation process.
The two beams are overlapped collinearly on a dichroic mirror and focused onto the HCPCF filled with Xenon gas (shown in Fig. \ref{fig:Mono_Sketch}), creating higher harmonics ($3\omega$ and above) via HCHG.
Typically, the driving pulses are focused to about \SI{50}{\micro\meter}, which is in correspondence to the HCPCF core diameter.
This results in a peak power of \SI{7}{\mega\watt} for each of the two driving pulses, corresponding to a peak power density of \SI{0.71}{\tera\watt\per{\centi\meter}^2}.

Depending on the phase matching pressure, HCHG results in high harmonics of the fundamental beam ($\omega = \SI{1.2}{\electronvolt}$) between \SIrange{5}{20}{\electronvolt}, with the harmonics separated by $\omega$, exceeding the ionization potential of Xenon \cite{couch_ultrafast_2020}.
Such harmonics are copropagating with the $1\omega$ and $2\omega$ components when leaving the fiber.
A Czerney-Turner type monochromator operating at near-normal incidence efficiently selects a single harmonic from \SIrange{7.2}{10.8}{\electronvolt} for measurements.
Fig. \ref{fig:Mono_Sketch} shows a sketch of the optical layout from the VUV generation to the monochromator's exit slit.
After the HCPCF, a retractable mirror can be inserted for alignment and diagnostic purposes.
The intensity of the $3\omega$ component, the lowest harmonic created in the HCHG process, is measured in this stage, using its power as a rough optimization parameter for temporal overlap between $\omega$ and $2\omega$.
The gas pressure inside the fiber and the delay are fine-tuned to optimize the relative yield of different higher-order harmonics.
The optimal phase matching pressure can vary slightly between different HCPCFs with the same properties (depending on manufacturing, degradation, etc.), as well as with the driving pulse properties and the desired VUV photon energy, but usually lies within \SIrange{100}{125}{\kilo\Pa}.
A thin film plate (TFP) with an anti-reflective (AR) coating for \SIlist{1030; 515; 357}{\nano\meter} at grazing incidence transmits and removes the most intense lower harmonics ($\omega$, $2\omega$ and $3\omega$), avoiding optical damage to the monochromator's optical components, while reflecting any VUV radiation.
For this purpose, all optics following the TFP and \ce{Au}-coated rejector pair are \ce{Al}-optics with a \ce{MgF2} capping layer.
The remaining higher harmonics are collimated by a curved mirror ($f=\SI{70}{\centi\meter}$) and diffracted from a grating with \SI{180}{grooves\per\milli\meter} at an angle of \SI{7.6}{\degree} to the surface normal.
Gratings with higher groove density could improve the energy resolution but at the cost of reduced photon flux and longer pulse duration.
A second curved mirror ($f=\SI{100}{\centi\meter}$) focuses the beam on a mask containing a set of four exit slits of \SIlist{3000; 500; 100; 32}{\micro\meter} in the horizontal dimension.
All slits have a vertical dimension of \SI{300}{\micro\meter} (see Fig. \ref{fig:Mono_Sketch}).

The temporal cross correlation between the VUV probe pulses and the pump pulse, measured at the sample's position determines the temporal resolution in the trARPES experiment.
In high-order harmonic generation processes the pulse duration of the emitted VUV light is typically shorter than the driving pulse.
However, the case of HCHG is more complicated, due to the interaction of multiple harmonic pulses co-propagating inside the \si{\cm}-long hollow core fiber, traveling through a partly ionized gas\cite{couch_ultrafast_2020}.
Generally speaking, the pulse duration and its structure are still an open question, and a first measurement of the limits of the pulse duration is provided in this report.
Additionally, monochromators using single-pass gratings increase the effective pulse duration by an induced pulse front tilt due to the diffraction from the grating\cite{martinez_pulse_1986,hebling_derivation_1996,poletto_time-preserving_2010,pennacchio_spatio-temporal_2018}.
This pulse front tilt is a direct consequence of the different path lengths for light scattered from different groves and is, therefore, proportional to the number of illuminated grooves on the grating.
Furthermore, the monochromated beam is relayed by a spherical mirror to the sample position, making the pulse front tilt proportional to the magnification factor of the imaging system between the grating and the sample.
From the parameters of our system, a pulse front tilt of \SI{430}{\femto\s} over the whole spot size can be estimated.
Reducing the monochromator's slit size reduces the time spread across the spot size and, therefore, the time resolution measured in a pump-probe experiment.

The last part of the laser is a separate diagnostics chamber, which is situated behind the exit slit and is used to monitor the photon flux with a retractable photodiode.
The diode has an \ce{Al2O3} coating and only responds to light of wavelngths smaller than \SI{200}{\nano\meter}, ignoring the possible stray light of the $1\omega$-$3\omega$ components.
Harmonic spectra are collected by rotating the monochromator grating using a motorized mount, while measuring the photon flux on the photodiode operating in photoconductive mode.
Here, the photoelectrons, generated by the VUV radiation, are detected as an electric current on a biased steel anode ($U_{Bias}=\SI{72}{\volt}$).
This current is then related to \si{\pps} via the wavelength-dependent responsivity of the photodiode.
The flux of the \SI{10.8}{\electronvolt} beam after the exit slit is up to \SI{1.5e11}{\pps} and can be maintained for several hours.
The system is typically operated at a lower flux, prioritizing longevity and providing a stable flux of $\sim\SI{3e10}{\pps}$ that can be maintained for long periods, with the HCPCF lasting for multiple months before any notable degradation.
The three other harmonics ($6\omega$, $7\omega$ and $8\omega$) typically show a flux that is one order of magnitude higher for each of them with approximately \SI{1e13}{\pps}, \SI{2e12}{\pps} and \SI{2e11}{\pps}, respectively.
The values of the flux for the different conditions are given for an exit slit width of \SI{500}{\micro\meter}.
Closing the slit from \SIrange{500}{100}{\micro\meter} reduces the flux roughly by a factor of 3, and further reduction from \SIrange{100}{32}{\micro\meter} reduces the flux again by a factor of approximately 3.
The photon flux delivered to the sample for all harmonics and slit combinations can be found in Tab. \ref{tab:e_res}.

\section{\lowercase{tr}ARPES setup}

\begin{figure}[b]
\includegraphics[width=\columnwidth]{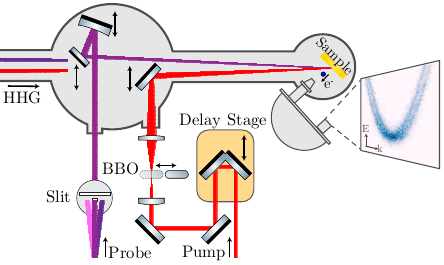}%
\caption{The trARPES setup includes the pump-probe beamline of the VUV laser source and the integration with the HHG source ("Harmonium").
\label{fig:System_Sketch}}
\end{figure}

The HCHG source is integrated into a preexisting HHG beamline ("Harmonium"\cite{ojeda_harmonium_2015}).
Harmonium provides light pulses in the EUV regime between \SIrange{20}{110}{\electronvolt} and achieves a time duration of <\SI{100}{\femto\s} at an energy resolution of $\sim\SI{150}{\milli\electronvolt}$ in the \SIrange{15}{40}{\electronvolt} range.
A circular vacuum chamber was inserted into the HHG beam path\cite{arrell_harmonium_2017} to deliver the VUV and HHG radiation to the ARPES chamber.
The chamber hosts a curved normal incidence \ce{MgF2}-coated \ce{Al} mirror ($f=\SI{100}{\centi\meter}$), placed at a small angle, to image-relay the divergent VUV beam and a motorized, flat \ce{MgF2}-coated \ce{Al} mirror to redirect the beam onto the sample.
The coating is optimized for \SI{120}{\nano\meter} and the two coated mirrors have each a reflectivity of \SIrange{78}{80}{\%}, which reduces the flux of the $9\omega$ beam delivered to the sample from \SI{3e10}{\pps} as measured by the photodiode to \SI{1.8e10}{\pps} (\SI{1.2e11}{\pps}, \SI{1.2e12}{\pps} and \SI{0.6e13}{\pps} delivered to sample for $8\omega$, $7\omega$ and $6\omega$).
All optics are installed on motorized stages, allowing for quick selection between the two light sources (see Fig. \ref{fig:System_Sketch}).
Additionally, the spot size on the sample can be optimized by adjusting the position of the vacuum stage hosting the curved mirror.
The spot size was estimated by measuring the zero-order $2\omega$ beam on a beam profiler situated outside the vacuum chamber, placed at the same distance as the sample.
This way a spot size of $\Delta x_{FWHM}=\SI{360}{\micro\meter} \times \Delta y_{FWHM}=\SI{220}{\micro\meter}$ for a slit width of \SI{500}{\micro\meter} and $\Delta x_{FWHM}=\SI{205}{\micro\meter}\times\Delta y_{FWHM}=\SI{220}{\micro\meter}$ for a slit width of \SI{100}{\micro\meter} can be observed. 
Due to the lower divergence of VUV light, this is expected to be an upper estimate of the experimental spot size.

The combination of both light sources in a single trARPES setup allows for switching between high energy and reasonably good time resolution at high repetition rates, and high time resolution at lower energy resolution at low repetition rates, depending on the characteristics of the observed effect.
It also offers the possibility to tune the photon energy in a large region and select a harmonic depending on the matrix element of a given sample, as well as providing selectivity regarding the region of interest in k-space.

Both beamlines have access to an IR pump and the possibility of frequency conversion via second harmonic generation (SHG) (Y-Fi: \SIlist{1.2; 2.4}{\electronvolt}; Harmonium: \SIlist{1.55; 3.1}{\electronvolt}).
The residual $\omega$ component of the HCHG source, which is not used for the VUV creation, exits the source from a second port.
A motorized attenuator regulates the pump fluence.
A delay stage is situated behind the attenuator to control the relative pump-probe delay, and a telescope consisting of two plano-convex lenses is used to adjust the focal spot on the sample.
The pump spot size was measured to be $\Delta x_{FWHM}=\SI{390}{\micro\meter}\times\Delta y_{FWHM}=\SI{360}{\micro\meter}$, uniformly covering the VUV spot.
Additionally, a BBO can be inserted into a focal spot between the two lenses, converting the pump energy from \SI{1.2}{\electronvolt} to \SI{2.4}{\electronvolt}.
After entering the circular chamber, a flat mirror can be inserted into the pump beam path with a vacuum translation stage, steering the beam toward the sample.
Figure \ref{fig:System_Sketch} provides a sketch of the trARPES setup from the VUV laser to the sample.

\section{Energy resolution measured on A\lowercase{u}(111), B\lowercase{i}\textsubscript{2}S\lowercase{e}\textsubscript{3} and \lowercase{polycrystalline} A\lowercase{u}}
\label{sec:energy}

Photoelectron spectra are collected with a hemispherical electron energy analyzer (Phoibos 150), capable of a resolution of \SI{5}{\milli\electronvolt}.
All samples are mounted on a home-built cryogenic manipulator.
Data were collected on polycrystalline \ce{Au}, single crystal \ce{Au(111)} and \ce{Bi2Se3} to assess the performance of the new light source at a repetition rate of \SI{1}{\mega\hertz}.

A systematic study of the total energy resolution and the laser's spectral bandwidth was done by measuring spectra of polycrystalline \ce{Au} for all possible harmonics and monochromator exit slit combinations.
A LHe cryostat was used to reduce thermal broadening with the spectra taken at a temperature $<\SI{20}{\kelvin}$.
In ARPES, the total energy resolution is a consequence of the material's intrinsic spectral width, broadened by the probe pulse spectral bandwidth and by the detection system's energy resolution.
Therefore, to evaluate the spectral bandwidth of the probe pulse, it is necessary to separate it from the other contributions.
This is achieved by measuring the same spectra under the same conditions with a Helium lamp, with a spectral bandwidth of a few \si{\milli\electronvolt}, which is a negligible contribution to the combined energy resolution and thereby corresponds to the detection system broadening, noted as $\Delta E_{He}$.
The total energy resolution is determined by plotting the EDC integrated over the whole momentum range and fitting a linear density of states multiplied with a Fermi-Dirac distribution and convolved with a Gaussian broadening.
This results in the following fitting function
\begin{equation}
    I = \left[ c +  \frac{1}{e^{(E-E_F)/k_BT}+1} \right] * g(E, \Delta E)
\label{eq:energy_res}
\end{equation}
with a constant background $c$, the kinetic energy of detected electrons relative to the Fermi level $E-E_F$, Boltzmann constant $k_B$, temperature $T$, and $*$ indicating the convolution operation.

\begin{figure}[b]
\includegraphics[width=\columnwidth]{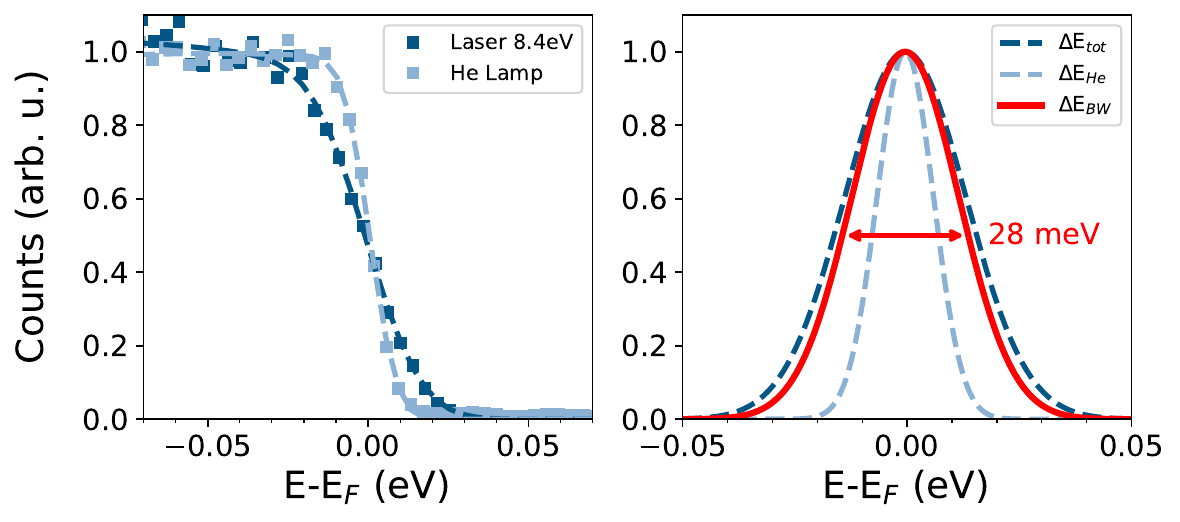}
\caption{Left: Two EDCs of polycrystalline Au measured in the same conditions. One using the VUV laser at \SI{8.4}{\electronvolt}, the other the Helium Lamp. A narrower Fermi-Edge is observed for the EDC obtained with the He Lamp. Right: Two Gaussians (dashed) are plotted with parameters extracted from the Fermi-Dirac fit of the EDCs on the left, representing the energy resolutions $\Delta E_{tot}$ and $\Delta E_{He}$. The third Gaussian (red, solid) is a result of the deconvolution, representing the spectral bandwidth $\Delta E_{BW}$ of the VUV harmonic.
\label{fig:Deconvolution}}
\end{figure}

The sample's temperature was fixed at a value of \SI{12}{\kelvin}, as measured by a thermal probe close to the sample, accounting for the thermal broadening and with that the material's intrinsic spectral width.

Deconvolving $\Delta E_{He}$, from the total energy resolution $\Delta E_{tot}$ measured with the VUV laser, reveals the spectral bandwidth of the VUV harmonic $\Delta E_{BW}$.
Fig. \ref{fig:Deconvolution} visualizes the deconvolution by comparing two EDCs, one measured with the He lamp and the other with the VUV laser's $7\omega$ harmonic (\SI{8.4}{\electronvolt}).
The Gaussian broadening, extracted from the Fermi-Dirac fits of Eq. \ref{eq:energy_res}, is shown in the right part of Fig. \ref{fig:Deconvolution} comparing $\Delta E_{tot}$ and $\Delta E_{He}$.
Additionally, the Gaussian of the calculated spectral bandwidth $\Delta E_{BW}$ is shown, revealing that the energy resolution of the system is dominated by the contribution of the VUV laser's spectral bandwidth.
All energy resolutions and spectral bandwidths are summarized in Tab. \ref{tab:e_res}.

\begin{table}[t]
    \renewcommand{\arraystretch}{1.2}%
    \caption{Deconvolved spectral bandwidths $\Delta E_{BW}$ and total energy resolution $\Delta E_{tot}$ for the different harmonics and monochromator exit slits with their corresponding photon flux delivered to the sample. Additionally, the energy resolution measured with the He lamp $\Delta E_{He}$, used for the deconvolution. An example of the deconvolved contributions is shown in Fig. \ref{fig:Deconvolution}.}
    \begin{center}
        {\setlength\doublerulesep{0.6pt}
        \begin{tabularx}{\columnwidth}{XXXXXX}
            \toprule[1pt]\midrule[0.3pt]
            $\hbar\omega$ (\si{\electronvolt}) & Slit width (\si{\micro\meter})  & $\Delta E_{BW}$ (\si{\milli\electronvolt}) & $\Delta E_{He}$ (\si{\milli\electronvolt})   & $\Delta E_{tot}$ (\si{\milli\electronvolt}) &  Flux (\si{\pps}) \\
            \midrule
            10.8    &    500    &    40      &    26     &      48      &       \num{1.8e10} \\
            10.8    &    100    &    25.5    &    26     &      36      &       \num{0.6e10} \\
            10.8    &    32     &    25.8    &    18     &      31.5    &       \num{0.2e10} \\
            9.6     &    500    &    44.5    &    18     &      48      &       \num{1.2e11} \\
            9.6     &    100    &    30      &    18     &      35      &       \num{0.4e11} \\
            9.6     &    32     &    20.4    &    14.5   &      25      &       \num{0.1e11} \\
            8.4     &    500    &    39.1    &    14.5   &      41.7    &       \num{1.2e12} \\
            8.4     &    100    &    26.5    &    13.8   &      29.9    &       \num{0.4e12} \\
            8.4     &    32     &    16.9    &    13.8   &      21.8    &       \num{0.1e12} \\
            7.2     &    500    &    38.7    &    17     &      42.2    &       \num{6.0e12} \\
            7.2     &    100    &    25.6    &    9.1    &      27.2    &       \num{2.0e12} \\
            7.2     &    32     &    18.3    &    9.1    &      20.4    &       \num{0.7e12} \\
            \midrule[0.3pt]\bottomrule[1pt]
        \end{tabularx}
        }
        \label{tab:e_res}
    \end{center}
\end{table}

\begin{figure}[t]
\subfloat[]{\includegraphics[width = \columnwidth]{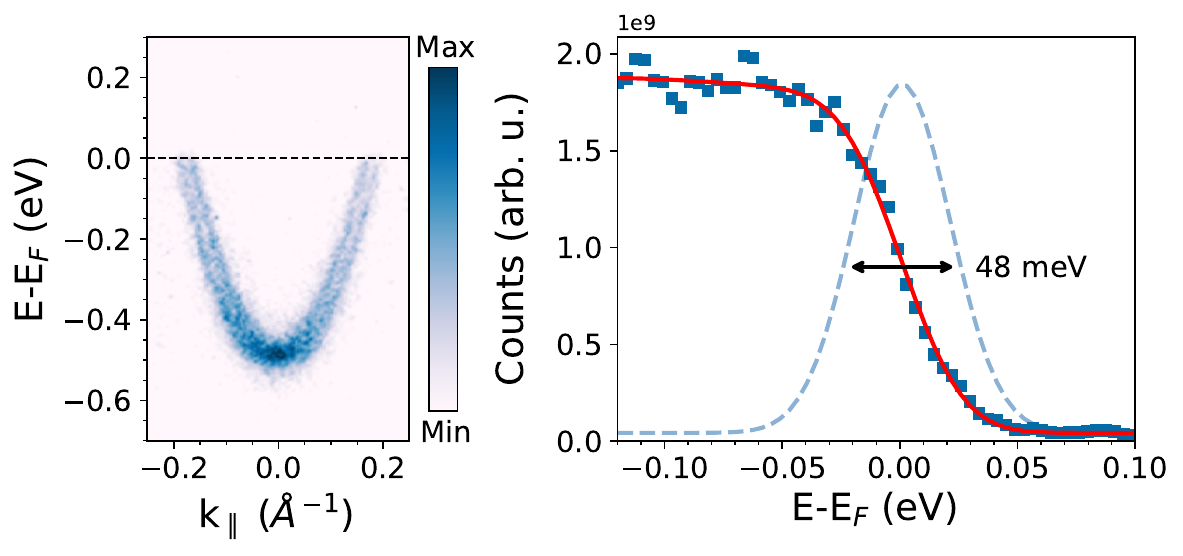}}\\
\subfloat[]{\includegraphics[width = \columnwidth]{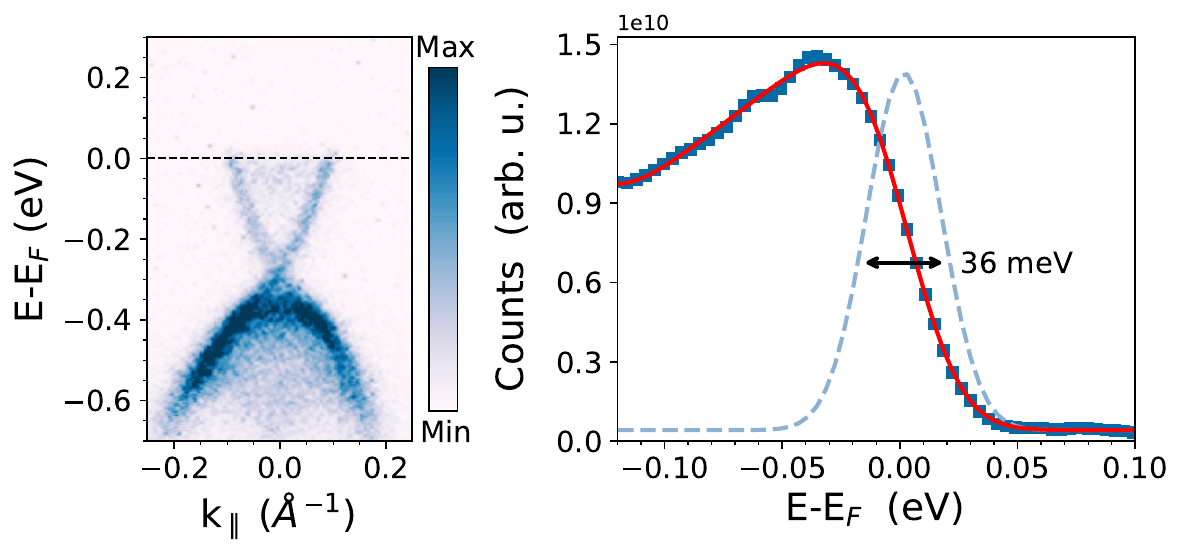}}
\caption{(a) Left: Cut of \ce{Au(111)} through the $\Gamma$-Point, showing the Rashba-split surface state. Right: Momentum integrated EDC, taken at higher momenta. Therefore, the integration is done over the background of elastically scattered electrons. Blue squares represent the measured intensity, the fitted curve according to Eq. \ref{eq:energy_res} in red, and the light blue dotted line represents the Gaussian broadening $g(E,\Delta E)$. Measurement conditions: $\hbar\omega=\SI{10.8}{\electronvolt}$, slit $=\SI{500}{\micro\meter}$, T $=\SI{12}{\kelvin}$.
(b) Left: Band map of \ce{Bi2Se3} intersecting the $\Gamma$-Point, showing the Dirac cone of the topological insulator. Right: Momentum-integrated EDC, with measured intensity (blue squares) and fit (red solid line) and the Gaussian broadening (light blue dotted line). Measurement conditions: $\hbar\omega=\SI{10.8}{\electronvolt}$, slit $= \SI{100}{\micro\meter}$, T $=\SI{65}{\kelvin}$.
\label{fig:EDC}}
\end{figure}

Two more samples, \ce{Au(111)} and \ce{Bi2Se3}, have been studied to characterize the system in a more realistic scenario.
\ce{Au(111)} was chosen as it provides a high-quality surface which was prepared by multiple cycles of \ce{Ar} ion sputtering and annealing to \SI{500}{K}, showing a clean Fermi-edge at high momenta.
Additionally, the light source makes the observation of well-defined, Rashba-type spin split surface states possible (see Fig. \ref{fig:EDC} (a)) without the need for additional optics\cite{kawaguchi_time-_2023}.
This measurement was performed with a photon energy of \SI{10.8}{\electronvolt} with a liquid Helium (\ce{LHe}) cryostat lowering the sample temperature to $<\SI{20}{\kelvin}$, reducing thermal broadening.

Energy distribution curves (EDC) were collected at a $k_\parallel$ away from any surface state that could influence the Fermi-Dirac distribution, to determine the energy resolution.
For these conditions a combined energy resolution of \SI{48}{\milli\electronvolt} at full-width half maximum (FWHM) was observed for the monochromator's \SI{500}{\micro\meter} wide exit.
Figure \ref{fig:EDC} (a) shows the integrated EDC together with the fit function.
Repeating the procedure to extract the spectral bandwidth described before yields a system energy broadening of $\Delta E_{He}=\SI{26}{\milli\electronvolt}$, resulting in a spectral bandwidth of $\Delta E_{BW}=\SI{40}{\milli\electronvolt}$ for the VUV laser's \SI{500}{\micro\meter} slit at an energy of \SI{10.8}{\electronvolt}.

We repeated the same measurement on \ce{Bi2Se3}, a well-studied topological insulator with a Dirac cone crossing the Fermi level.
This Dirac cone also provides a clean edge, well suited for estimating the energy resolution.
The higher photo-electron yield also produces a better signal-to-noise ratio, enabling a more precise estimate for smaller monochromator slits at a photon energy of \SI{10.8}{\electronvolt}.
The sample was cleaved in situ, and measurements were performed at low temperatures using a liquid nitrogen (LN) cryostat.
The bandmap was measured as a cut through the $\Gamma$-Point, and the momentum-integrated EDC takes the Dirac cone region into account (see Fig. \ref{fig:EDC} (b)).
From the fit, a combined energy resolution of \SI{36}{\milli\electronvolt} can be extracted for the monochromator slit with a width of \SI{100}{\micro\meter}.
Deconvolving the total energy resolution $\Delta E_{tot}$ and detection system broadening $\Delta E_{He}$ yields a spectral laser bandwidth as low as \SI{25.5}{\milli\electronvolt}.

In total, an ultimate energy resolution of \SI{21}{\milli\electronvolt} was found for a photon energy of \SI{7.2}{\electronvolt} in combination with a monochromator exit slit of \SI{32}{\micro\meter}.
While the usage of narrower slits improves the energy resolution it comes at a cost of a reduced photon flux.
The combined energy resolution of the measurement setup and the spectral bandwidth for the different harmonics and differently sized slits are summarized in Table \ref{tab:e_res}, together with their corresponding photon flux on the sample.

\section{Time-resolved measurements on B\lowercase{i}\textsubscript{2}S\lowercase{e}\textsubscript{3} and T\lowercase{a}T\lowercase{e}\textsubscript{2}}

We characterized the time resolution of the measurement setup by measuring the band-filling process above $E_F$ and its subsequent decay on both \ce{Bi2Se3} and \ce{TaTe2}.
As described in the section above, \ce{Bi2Se3} hosts a Dirac cone crossing the Fermi level, which can be occupied above $E_F$ upon pump excitation and is, therefore, a good candidate for evaluating the performance of the trARPES setup.
The sample was cleaved in situ, and measurements were carried out at room temperature.
The 10\textsuperscript{th} harmonic (\SI{10.8}{\electronvolt}) was chosen as a probe beam for the pump-probe measurements, while the \SI{1.2}{\electronvolt} fundamental beam was used to pump the material.
Both samples were investigated with a repetition rate of \SI{1}{\mega\hertz}, similar to the previous section.
The slit with a width of \SI{100}{\micro\meter}  (see Fig.\ref{fig:Mono_Sketch}) was used for the experiment for better temporal pump-probe cross-correlation.

\begin{figure*}[t]
\includegraphics[width =2\columnwidth]{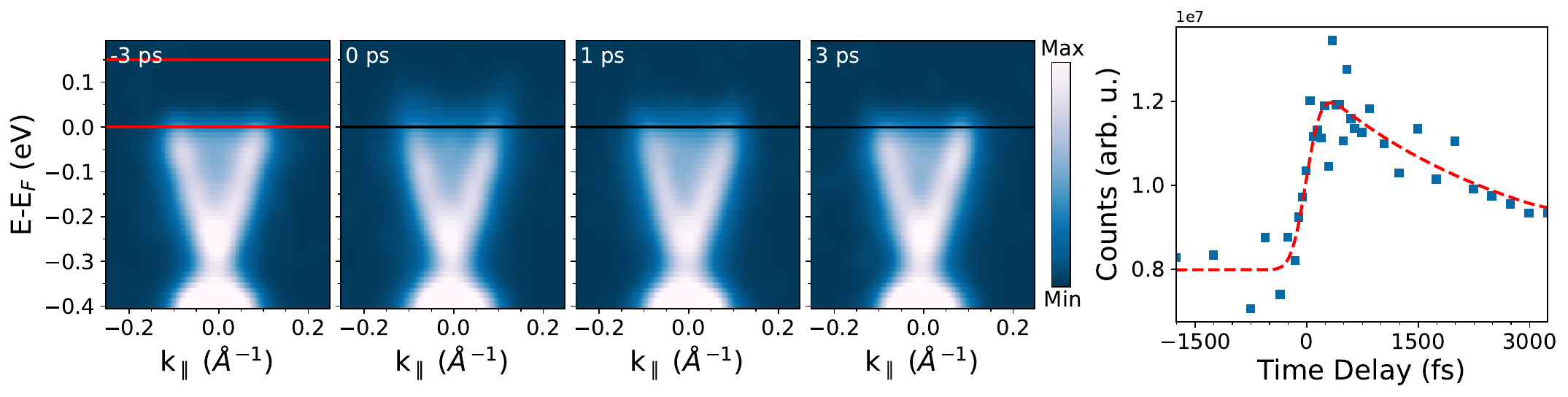}
\caption{Series of cuts intersecting the Dirac cone for different time delays, showing the fast population above $E_F$ and the subsequent decay. Two horizontal red lines in the first delay step mark the integrated area displayed as a function of time delay on the right. Blue squares represent the measured intensity for different delay steps together with the fitted decay model (red dotted curve) according to Eq. \ref{eq:time_res}. Measurement conditions: $\hbar\omega=\SI{10.8}{\electronvolt}$, slit $= \SI{100}{\micro\meter}$, T $=\SI{65}{\kelvin}$.
\label{fig:bise_dyn}}
\end{figure*}

We observe a fast band filling of the unoccupied part of the Dirac cone, followed by slow relaxation on the order of picoseconds, similar to previous reports\cite{wang_measurement_2012, crepaldi_ultrafast_2012}.
Figure \ref{fig:bise_dyn} shows cuts of the Dirac cone filling for four different time delays.
Since the population of the Dirac cone occurs on a short \si{\femto\s} timescale, the observed rise time is dominated by the laser pulse duration.
To extract the time resolution, an exponential decay with a step at time-zero, convoluted with a Gaussian is fitted to the data, as given by the expression
\begin{equation}
   I = \left[ c + A  \theta(t_0)  e^{\left(\frac{-(t-t_0)}{\tau} \right) }  \right] * g(t, \Delta t)
\label{eq:time_res}
\end{equation}
with $c$ as constant background, $A$ as amplitude, $\theta$ as a Heavyside function centered at $t_0$, $\tau$ as decay rate, and $*$ indicates the convolution operation.
The width of the Gaussian $g(t, \Delta t)$ represents the temporal instrument response function.
The resulting time trace is plotted with the fit function in Fig. \ref{fig:bise_dyn}.
A time resolution of \SI{360}{\femto\s} as the FWHM of the Gaussian can be extracted for a slit-width of \SI{100}{\micro\meter}.

Additionally, we performed pump-probe measurements on \ce{TaTe2}, a metallic system hosting multiple CDWs, with the room temperature (RT) ground state exhibiting a commensurate CDW (CCDW)\cite{feng_charge_2016, chen_trimer_2018}.
\ce{TaTe2} was measured to further asses the performance in terms of time resolution.
As in the case of \ce{Bi2Se3}, the samples have been cleaved in situ, and measurements have been performed at a low temperature of \SI{65}{\kelvin} with the help of a \ce{LN2} cryostat.
Since \ce{TaTe2} undergoes a structural phase transition at $T \sim \SI{170}{\kelvin}$, the measurement was performed in the incommensurate CDW (ICDW) or "CDW-like" state, as referred to in the literature\cite{gao_origin_2018, chen_trimer_2018, lin_evidence_2022}.
However, the exact mechanism of phase transition and the formation of the CDW are still under debate.
trARPES and other ultrafast techniques could help develop a better understanding of the materials' different phases.
Some work with other methods has been carried out\cite{siddiqui_ultrafast_2021}, but to our knowledge, there is currently no trARPES data published on this compound, and we present novel data.
A full description of the dynamics would go beyond the scope of this report, but the experimental setup described in this report will be able to address these questions in the future.
A monochromator slit of \SI{500}{\micro\meter} was used for the measurements.

\begin{figure}[h!]
\includegraphics[width = \columnwidth]{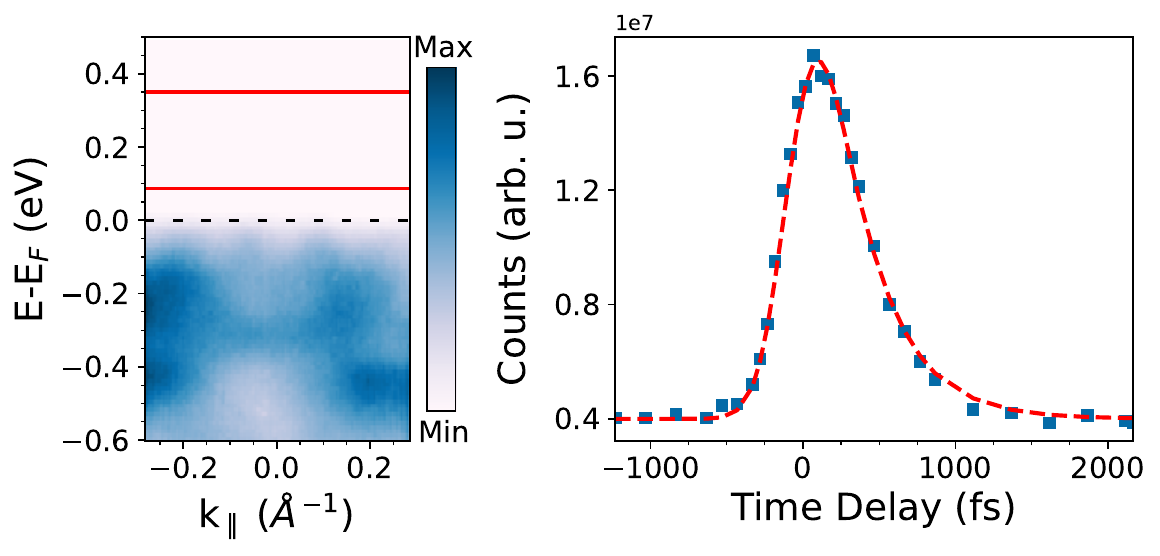}
\caption{Left: EDC of \ce{TaTe2}. The red area marks the unoccupied region used to extract the time resolution from the time trace on the right. Right: Dynamics of the unoccupied region of \ce{TaTe2} after pump excitation (blue), fitted with an exponential decay model according to Eq. \ref{eq:time_res} (red dashed line). The fit shows a time resolution of \SI{410}{\femto\s}. Measurement conditions: $\hbar\omega=\SI{10.8}{\electronvolt}$, slit $= \SI{500}{\micro\meter}$, T $=$ RT.
\label{fig:tate_dyn}}
\end{figure}

The time evolution of the unoccupied region \SI{100}{\milli\electronvolt} above $E_F$ has an intrinsically fast rise time dominated by the probe pulse time duration.
The band filling is followed by a rapid decay, with the population fully relaxing within the first \SI{1.5}{\pico\s}.
The time-evolution is fitted again by the convolution of an exponential decay and step function with a Gaussian, described by Equation \ref{eq:time_res}.
From this, a time resolution of \SI{410}{\femto\s} is extracted.
Figure \ref{fig:tate_dyn} shows a cut of the \ce{TaTe2} band structure and the time evolution of the integrated area above $E_F$ and the fit, corresponding to Equation \ref{eq:time_res}. 

These results confirm that the pulse front tilt is the dominating contribution to the time resolution, since the extracted value of \SI{410}{\femto\s} is of the same order as the estimated pulse front tilt due to the monochromator.
The pulse front tilt is reduced by closing the exit slit and clipping the beam.
Hence, the time resolution is enhanced at the cost of a lower photon flux, which can be seen in the trace recorded using the \SI{100}{\micro\meter} wide slit.
The resulting time resolutions of the different monochromator exit slits are summarized in Table \ref{tab:t_res}, together with their corresponding photon flux as measured after the monochromator exit slit.

\begin{table}[h!]
    \renewcommand{\arraystretch}{1.2}%
    \caption{Time resolution for the differently sized monochromator exit slits and their corresponding photon flux delivered to the sample. Measurements were taken at a photon energy of \SI{10.8}{\electronvolt}.}
    \begin{center}
        {\setlength\doublerulesep{0.6pt}
        \begin{tabularx}{\columnwidth}{XXXX}
            \toprule[1pt]\midrule[0.3pt]
            $\hbar\omega$ (\si{\electronvolt}) & Slit width (\si{\micro\meter})  & $\Delta t$ (\si{\femto\s})  &  Flux (\si{\pps})       \\
            \midrule
            10.8        &       500         &       410      &     \num{1.8e10}        \\
            10.8        &       100         &       360      &     \num{0.6e10}        \\
            \midrule[0.3pt]\bottomrule[1pt]
        \end{tabularx}
        }
        \label{tab:t_res}
    \end{center}
\end{table}

\section{Conclusion}

We presented a measurement system consisting of a new state-of-the-art VUV light source and an HHG extreme ultraviolet (EUV) source combined, performing trARPES experiments.
The VUV light source operates at high repetition rates of \SIrange{0.5}{2}{\mega\hertz}, significantly reducing space charge issues, while also suffering less from sample heating.
A spectral bandwidth of \SI{17}{\milli\electronvolt} was observed while maintaining a time resolution of <\SI{360}{\femto\s}.
The pulse-front tilt induced by the monochromator's grating mainly limits the time resolution.
However, our experiments indicate that a near-NI monochromator layout is sufficient for trARPES experiments at high energy resolution in the VUV regime.
The achieved setup resolution of $\Delta E=\SI{21}{\milli\electronvolt}$ and $\Delta t=\SI{360}{\femto\s}$ is equal or slightly worse than conventional third harmonics-based setups\cite{peli_time-resolved_2020, lee_high_2020, kawaguchi_time-_2023}.
A way to improve the time resolution of our setup could be the introduction of a double grating monochromator, compensating the pulse front tilt of the probe pulse and sacrificing little in energy resolution.
Additionally, the HCHG process used to create the higher harmonics offers tunability between \SIrange{7.2}{10.8}{\electronvolt}, enabling larger energy and momentum selectivity, which is not possible in third harmonics-based setups.

The VUV light source is incorporated into the existing HHG beamline, extending the tunability to the EUV regime between \SIrange{20}{110}{\electronvolt} at low repetition rates from \SIrange{2}{6}{\kilo\hertz}.
The HHG beamline offers the advantage of a higher temporal resolution of <\SI{100}{\femto\s} at a reduced energy resolution of <\SI{150}{\milli\electronvolt}.
Together, these two complementary beamlines allow for extreme flexibility to tackle new complex materials and conduct comprehensive studies in a lab environment.
The trARPES setup described in this article can be tuned in an extreme photon energy range, from the less accessible VUV to the deep EUV regime.
Additionally, it allows both the observation of very fast phenomena and the dynamics of lower energy features, such as high-temperature superconducting gap throughout the FBZ.

\begin{acknowledgments}
We acknowledge financing through the Lausanne Center for Ultrafast Science (LACUS), as well as support by the European Research Council through the Consolidator network ISCQuM (Grant No. 771346) and the Synergy network HERO (Grant No. 810451).
\end{acknowledgments}

\section*{Data Availability Statement}
The data that support the findings of this study are openly available in the Dataset of "High-resolution MHz trARPES based on a tunable VUV source" at http://doi.org/10.5281/zenodo.8386933, reference number 8386933.

\bibliographystyle{ieeetr}
\bibliography{VUV_bib_2}

\end{document}